\documentclass[aps,prl,twocolumn,showpacs,nofootinbib]{revtex4-1}
\usepackage{amsmath,amssymb,graphicx}
\usepackage{epsf}
\usepackage{bm}

\newcommand{\nc}{\newcommand}
\nc{\postscript}[2]
{\setlength{\epsfxsize}{#2\hsize}\centerline{\epsfbox{#1}}}
\nc{\non}{\nonumber}
\nc{\hc}{\hbox {h.c.}} \nc{\re}{\hbox {Re}} 
\nc{\mev}{\hbox {MeV}} \nc{\gev}{\;\hbox {GeV}} \nc{\tev}{\;\hbox {TeV}}
\def\lsim{\mathrel{\raise.3ex\hbox{$<$\kern-.75em\lower1ex\hbox{$\sim$}}}}
\def\gsim{\mathrel{\raise.3ex\hbox{$>$\kern-.75em\lower1ex\hbox{$\sim$}}}}

\nc{\etal}{{\it et al.}}
\nc{\Lsp}{\;\;\;\;\;\;\;\;\;\;}  \nc{\LLLsp}{\lspace \lspace}
\nc{\lsp}{\;\;\;\;\;\;}
\nc{\spac}{\;\;\;}
\nc{\noi}{\noindent}
\nc{\beq}{\begin{equation}}   \nc{\eeq}{\end{equation}}
\nc{\bea}{\begin{eqnarray}}   \nc{\eea}{\end{eqnarray}}
\nc{\baa}{\begin{array}}      \nc{\eaa}{\end{array}}
\nc{\bit}{\begin{itemize}}    \nc{\eit}{\end{itemize}}
\nc{\ben}{\begin{enumerate}}  \nc{\een}{\end{enumerate}}
\nc{\bce}{\begin{center}}     \nc{\ece}{\end{center}}

\def\sq2{\sqrt{2}}

\def\ph{\varphi}

\def\m4{m^4(\ph)}
\def\mn2{m_n^2}

\def\v5{V^{(5)}}

\begin{document}

\title{\begin{flushright}
       \mbox{\normalsize \rm CUMQ/HEP 182,~
      \rm  UQAM-PHE/02-2014}
       \end{flushright}
       \vskip 15pt
Unified Flavor Symmetry from warped dimensions
}
\author{Mariana Frank$^{(1)}$\footnote{Email: mariana.frank@concordia.ca}}
\author{Cherif Hamzaoui$^{(2)}$\footnote{Email: xx}}
\author{Nima Pourtolami$^{(1)}$\footnote{Email: n\_pour@live.concordia.ca}}
\author{Manuel Toharia$^{(1)}$\footnote{Email: mtoharia@physics.concordia.ca}}
\affiliation{$^{(1)}$Department of Physics, Concordia University,
7141 Sherbrooke St. West, Montreal, Quebec,\ Canada, H4B 1R6,}
\affiliation{$^{(2)}$Groupe de Physique Th\'eorique des Particules,
D\'epartement des Sciences de la Terre et \\de L'Atmosph\`ere,
Universit\'e du Qu\'ebec \`a Montr\'eal, Case Postale 8888, Succ. Centre-Ville,
 Montr\'eal, Qu\'ebec, Canada, H3C 3P8.}

\date{\today}

\begin{abstract}
We propose a scenario which accommodates all the masses and
mixings of the SM fermions in a model of warped extra-dimensions with all matter fields in
the bulk.  In this  
scenario, the same flavor symmetric structure is imposed on all
the fermions of the Standard Model (SM), including neutrinos.
Due to the exponential sensitivity on bulk fermion masses, a small
breaking of the symmetry can be greatly enhanced and produce
seemingly un-symmetric hierarchical masses and small mixing angles among the
charged fermion zero-modes (SM quarks and charged leptons) and
wash-out the obvious effects of the symmetry. 
With the Higgs field
leaking into the bulk, and Dirac neutrinos sufficiently localized towards the
UV boundary, the neutrino mass hierarchy and flavor structure will
still be largely dominated by the fundamental flavor structure.
The neutrino sector would then reflect  the fundamental
flavor structure, whereas the quark sector
would probe the effects of the flavor symmetry breaking sector.
As an example, we explore these features in the context of a family
permutation symmetry imposed in both quark and lepton sectors.

\end{abstract}

\maketitle

The original motivation for warped extra-dimensions, or Randall-Sundrum models (RS), was to
address the hierarchy problem. In RS the fundamental scale of gravity
is exponentially reduced from the Planck mass scale to a TeV
size due to a Higgs sector localized near the boundary of the extra dimension \cite{RS1}.
If SM fermions are allowed to propagate in the extra dimension \cite{Gherghetta:2000qt},
and become localized towards either boundary, the scenario also
addresses the flavor problem of the SM and suppresses generic
flavor-violating higher-order operators present in the original RS
setup. However, KK-mediated processes still generate dangerous
contributions to electroweak and flavor observables \cite{RSeff,Burdman,Agashe},
pushing the KK scale to 
$5-10$ TeV \cite{Weiler}.  
One realization of the model is based on the so-called flavor
anarchy \cite{Agashe}, in which one assumes that no special structure governs the
flavor of Yukawa couplings and bulk fermion masses, as natural
${\cal O}(1)$ values for these 5D parameters already generate viable
masses and mixings. The neutrino sector must behave
differently, first due to the possibility of Majorana mass terms, and
second because this setup generates large mass hierarchies and
small mixing angles, at odds with neutrino observations. 
An interesting property of 
warped scenarios was investigated
in \cite{Agashe:2008fe}, for the case of a bulk Higgs wave function leaking into the
extra dimension. There one would 
obtain small mixing angles and hierarchical masses for all charged fermions, and at the same
time very small Dirac masses, with large
mixing angles and negligible mass hierarchy for neutrinos.  Thus the flavor anarchy paradigm
could still work in these scenarios.

Here we present a scenario where instead of adopting flavor anarchy,
we propose that all fermions share the same flavor symmetry. We assume
that the flavor violating effects in the 5D Lagrangian can be
parametrized by a small coefficient whose size is controlled by a
ratio of scales,  $ {<\phi>^n \over \Lambda^n}$, with $<\phi>$ the
vacuum expectation value (VEV) of some flavon field, and $\Lambda$
some cut-off mass scale, or the KK mass of some other flavon fields.
This small breaking of the flavor symmetry is enough to
reproduce correctly the flavor structure of the SM in both the quark
and lepton sectors.
The (stable) static spacetime background is:
\bea
ds^2 = e^{-2A(y)}\eta_{\mu\nu} dx^\mu dx^\nu - dy^2,
\label{RS}
\eea
where the extra coordinate $y$ ranges between the
two boundaries at $y=0$ and $y=y_{\rm TeV}$, and where $A(y)$ is the warp factor
responsible for exponentially suppressing mass scales at different
slices of the extra dimension. In the original RS scenario $A(y)=ky$,
with $k$ the curvature scale of the $AdS_5$ 
interval, while in general warped scenarios $A(y)$ is a more general
(growing) function of $y$. The appeal of more complicated
metrics lies on the possibility of having light KK resonances ($\sim
1$ TeV), while keeping flavor and precision electroweak bounds at bay
\cite{Cabrer,Carmona}. For simplicity, we
assume $A(y)=ky$, unless otherwise specified.
Assuming invariance under the usual SM gauge group, the 5D quark Lagrangian
is
\begin{eqnarray}
{\cal L}_q&=& 
 {\cal L}_{kinetic} + M_{q_i}{\bar Q_i}Q_i + M_{u_i}{\bar U_i}U_i+ M_{d_i}{\bar D_i}D_i \non \\
&& + (Y^{5D}_{ij_u} H\bar{Q}_iU_j + h.c.) + (Y^{5D}_{ij_d} H\bar{Q}_iD_j + h.c.) \vphantom{\int} ,
\end{eqnarray}
where $Q_i$, $U_i$ and $D_i$ are 5D quarks (doublets and singlets
under $SU(2)$). In the lepton sector, we assume that Majorana mass
terms are forbidden, and so the Lagrangian can be trivially obtained
from the previous one by substituting $Q_i$ by $L_i$, $U_i$ by $N_i$ and $D_i$
by $E_i$, where $L_i$ are lepton doublets, and $N_i$ and
$E_i$ are neutrino and lepton singlets, respectively.
The Higgs field $H$ is a bulk scalar that acquires a nontrivial  VEV
$v(y) = v_0 e^{aky}$, exponentially localized towards the TeV
boundary, with delocalization controlled by the
parameter $a$. Such nontrivial exponential VEV-s appear naturally
in warped backgrounds with simple scalar potentials and appropriate
boundary conditions.
This extra dimensional scenario has two sources of flavor.
One is the usual Yukawa couplings $Y^u_{ij}$, $Y^d_{ij}$,
$Y^e_{ij}$ and $Y^\nu_{ij}$ (dimensionless parameters
defined in units of the curvature out the dimension-full 5D Yukawa
couplings as $Y_{ij}^{5D} = \sqrt{k} Y_{ij}$). The other 
comes from the fermion bulk mass terms, diagonal in flavor space,  taken to be constant bulk terms written in units of
the curvature $k$, i.e.  $M_i= c_i\ k  $ ($M_i=M_{q_i},
M_{u_i}, M_{d_i}, M_{L_i}, M_{\nu_i}, M_{e_i}$).

As noted in \cite{Agashe:2008fe},
whenever the bulk Higgs localization parameter $a$ is
small enough in comparison with the $c_i$ parameters, (i.e., for the Higgs
sufficiently delocalized from the TeV brane),
the 4D effective masses depend exponentially on $a$ rather than 
on the $c_i$ parameters.
The effective 4D masses for all the SM fermions become
\begin{eqnarray}
\label{top}m_{ t} &=&   v \tilde{Y}_{33} \hspace{3cm}
  c_{q_3} , c_{u_3} <1/2 \\
\label{fermions}(m_f)_{ij} &=&   v\epsilon^{(c_{L_i}-\frac{1}{2})}\epsilon^{(c_{R_j}-\frac{1}{2})}\tilde{Y}_{ij} \hspace{.7cm}
  a > c_{L_i}+c_{R_j}\\
\label{neutrinos}(m_\nu)_{ij}&=& v\epsilon^{a-1}\tilde{Y}_{ij} \hspace{2.45cm} a < c_{l_i}+c_{\nu_j} \, ,
\end{eqnarray}
where  $m_{ t}$ is the top quark mass, $(m_f)_{ij}$ are mass
matrices for light quarks and charged leptons, and  $(m_\nu)_{ij}$  is
the  Dirac neutrino mass matrix. The parameters $c_{L_i} \equiv
c_{q_i},c_{l_i}$ correspond to the $SU(2)$ doublets, and
$c_{R_j}\equiv c_{u_j},c_{d_j}, c_{e_j}, c_{\nu_j}$ are for the
$SU(2)$ singlets. 
The warp factor $\epsilon$ defined by the
background parameters as $\epsilon=e^{-k y_{TeV}}\sim 10^{-15}$ encapsulates
 the hierarchy between the UV (gravity) brane  and the TeV (SM) brane. 
\begin{figure}[t]
\vspace{-.2cm}
\center
\includegraphics[width=7.6cm]{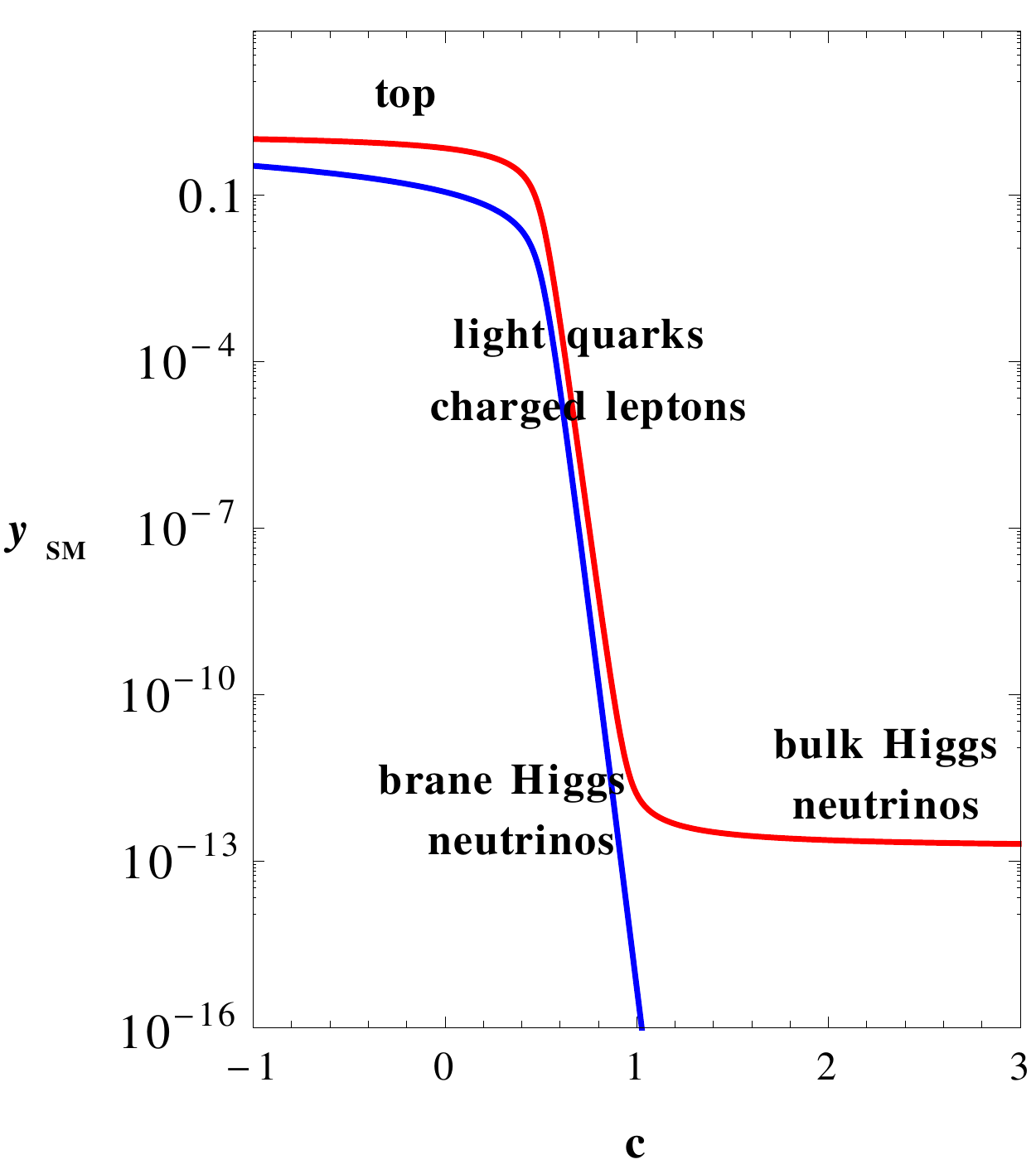}\ \ \ \ \
\vspace{-.4cm}
\caption{Effective 4D Yukawa couplings for fermions as a function of
  the fermion bulk mass parameter $c$. For simplicity, we have taken
  the $c$-parameters for the doublet and the singlet to be equal.
}
\label{plateau}
\end{figure}
The $c$-parameters dependence of masses is shown in Fig.
\ref{plateau} for the case of a brane localized Higgs VEV ($a=30$) 
and  for a delocalized Higgs VEV ($a=1.9$), where  the appearance of a
plateau in the neutrino mass region reflects the insensitivity to the  $c_i$ values in that limit. 
A source of tension arises since, in order to generate viable neutrino masses from equation
(\ref{neutrinos}), one requires that  $a\sim 1.80-1.95$. Values
of $a<2$ will reintroduce some amount of tuning in the
model and, for example, for $a=1.95$, some independent parameters of
the 5D Higgs potential must be fixed to be equal to within about $1\%$. However, this same tuning
will also be responsible for generating a light enough Higgs mode compared to the KK
scale \cite{Cabrer}. In more general warped backgrounds this tension
can easily disappear due to an enlarged parameter space, justifying
the choice $a=1.9$ throughout the rest of the paper.  

We assume that all  Yukawa matrices and fermion bulk
masses from the 5D Lagrangian share the same symmetry structure
broken by small terms i.e.
\bea
{\bf c}_f&=&{\bf c}_f^0 + \delta \bf{c}_f \label{cexpansion}\\
Y_{Y}&=&Y_{Y}^0 + \delta Y_{Y} \, , \label{Yexpansion}
\eea
for all fermions of the model with $Y_Y=Y_u,Y_d,Y_e,Y_\nu$ and ${\bf
  c}_f={\bf c}_q,{\bf c}_u,{\bf c}_d,{\bf c}_l,{\bf c}_e,{\bf c}_\nu$.
The matrices $Y^0_Y$ and ${\bf c}^0_f$  are flavor symmetric, while the
small corrections $\delta {\bf c}_f$ and $\delta Y_{Y}$
do not have {\it a priori} any flavor structure.

From Eqs. (\ref{top})-- (\ref{neutrinos}),
the fermion masses 
receive different corrections due to flavor violating
terms:
\bea
m_{t} &=& m^0_{t} +\delta m_{t}  \hspace{2cm} c_{q_3} , c_{u_3} <1/2 \\
(m_f)_{ij} &=& (m_f)^0_{ij}\  \epsilon^{( \delta c_{L_i}+ \delta c_{R_j})}  \hspace{.9cm}
  a > c_{L_i}+c_{R_j} \\
(m_\nu)_{ij}&=& (m_\nu)^0_{ij} + \delta (m_\nu)_{ij}  \hspace{1cm} a< c_{l_i}+c_{\nu_j}\, .
\eea
The exponential sensitivity on the $c$-parameters is
responsible for an exponential sensitivity of symmetry breaking
terms. Since $\epsilon \sim 10^{-15}$, the corrections to the mass
matrices caused by these are of order $\sim
10^{-15 ( \delta c_i+ \delta c_j)}$, which means they   
could account for the observed hierarchies in the quark
and charged lepton masses, as long as the symmetry breaking corrections $\delta c_i $
remain between $-0.1$ and $+0.1$.
The mixing angles are also  exponentially sensitive to small symmetry breaking terms so that the
mixing angles diagonalizing the mass matrices from the left  will be
$V_{ij} \sim  \epsilon^{(\delta c_{Li} - \delta c_{Lj})}$
for $(i < j) $ for quarks and charged leptons.  Thus effects of the
original symmetry are washed out in the quark and charged lepton sectors,
while in the Dirac neutrino sector the sensitivity to the symmetry
breaking is linear (i.e. small).  

To qualify our assertions, we impose a simple structure for all the flavor parameters
of the model, namely one which remains invariant under family
permutations \cite{s3}. This leads to a flavor structure where the 5D
Yukawa couplings are invariant under $S_3 \times S_3$, while the 5D
fermion bulk mass matrices are invariant under $S_3$.  This leads to democratic 5D Yukawa couplings and to 5D fermion bulk mass matrices parametrized as 
\bea
Y^D_{Y} \propto 
\left(\begin{array}{ccc}
1 & 1 & 1 \\
1 & 1 &  1\\
1 & 1 & 1\end{array}\right) \ \  {\rm and}\ \
{\bf c}^D_{f}=\!\! \left(\begin{array}{ccc}
 A_f & B_f & B_f \\
B_f &  A_f &  B_f\\
B_f & B_f &  A_f\end{array}\right)\,. \ \ \  \label{democracy}
\eea
Since all flavor structure is described by Eq.~(\ref{democracy}),  we
simultaneously diagonalize all matrices 
and obtain 
\bea
Y^{0}_{Y}= y^0_{{}_Y} \left(\begin{array}{ccc}
0 & 0 & 0 \\
0 & 0 &  0\\
0 & 0 & 1\end{array}\right) \  {\rm and}\ \
{\bf c}^{0}_{f}=\!\! \left(\begin{array}{ccc}
 {c^0_{f1}} & 0 & 0 \\
0 & c^0_{f1}  &  0\\
0 & 0 & c^0_{f3} \end{array}\right),\ \ \ \ \ \ \label{Ycdem}
\eea
where $y^0_{{}_Y}=y_{{}_u}, y_{{}_d}, y_{{}_e}, y_{{}_\nu}$ are complex
Yukawa couplings in the up, down, charged lepton and
neutrino Yukawa sectors. The matrix ${\bf c}_f$ is in its
diagonal basis with real entries and ${\bf c}_f^0$. Democratic mass matrices 
produce two massless fermions and one massive one. The 0-th order CKM
and PMNS matrices can be parametrized as 
\bea
V^0_{i}\!=
\left(\begin{array}{ccc}
\vphantom{\frac{9}{\sqrt{3}}}    \cos{\theta^0_{i}}   &
\sin{\theta^0_{i}}  &   \  0\  \\
\vphantom{\frac{\sqrt{9}}{\sqrt{3}}}    -\sin{\theta^0_{i}}  &  \cos{\theta^0_{i}}  &\   0\  \\
\vphantom{\frac{9}{\sqrt{3}}}         0       &     0        &   \ 1\
\end{array}\right)\, , \ \
\label{zeroorder}
\eea
where $i=$CKM, PMNS.  
Both matrices contain an angle
$\theta_i^0$, not fixed by the $S_3\times S_3$
symmetry, but by the symmetry
breaking terms, 
implemented  in Eqs.~(\ref{cexpansion}) and (\ref{Yexpansion}). The small breaking of the symmetry 
is responsible for a small lift of the zero masses,
yielding a viable neutrino spectrum,
with one heavier 
and two
lighter eigenstates with similar masses,
and a {\it normal} hierarchy ordering:
\bea
&m_3& \sim \frac{m^0_\nu}{  \sqrt{|\epsilon^{(2c_{l3}-1)} -1|}},\\ 
&m_1& \sim  \delta Y_\nu\ m^0_\nu \qquad m_2 \sim  \delta Y_\nu\ m^0_\nu\, ,
\eea
where $m^0_\nu=v\ \epsilon^{a-1}$. We have kept the term 
in $c_{l3}$ since it can have an effect when $c_{l3} \lsim \frac{1}{2}$. 
Neutrino mass data 
require $m^0_\nu \sim (0.05-0.1)$ eV, which in turn fixes the size
of the Higgs localization parameter $a$. 
The dependence on $\delta Y_\nu$ is evident while the $\delta c_i$'s
are basically free (even the 0-th order $c^0_{\nu i}$ are relatively free, as
long as they satisfy $a<c_\nu+c_l$).
For example, one finds that $\delta Y_\nu \lsim \sqrt{r}\sim 0.17$, to generate a viable neutrino mass hierarchy ratio
$r=(|m_2|^2-|m_1|^2)/(|m_3|^2-|m_1|^2)\sim 0.03$.

In the charged lepton, up-and down-quark sectors, the massless
states are also lifted by the flavor symmetry breaking leaving a suppression
proportional to $\delta Y$.  
In addition, the exponential dependence on the symmetry breaking parameters $\delta c_{f_i}$
creates a hierarchy among all the masses. The third generation charged
fermion masses are
\bea
m_t &\sim& m^0_{t}, \\
m_b & \sim& m^0_{b}\ \epsilon^{\delta c_{e_3}}, \\
m_\tau &\sim & m^0_{\tau}\ \epsilon^{(\delta c_{l_3}+ \delta c_{e_3})}\, ,
\eea
with the 0-th order masses\footnote{Assuming that $c^0_{q3},c_{u3}<\frac{1}{2}$ and
$c^0_{d3},c^0_{e3},c^0_{l3}>\frac{1}{2}$.} 
$\ m_t^0=y^0_u\ v\,$, $m_b^0 = y^0_d v\ \epsilon^{c^0_{d3}-1/2}\,$
and $\ m_\tau^0 =
y^0_e v\ \epsilon^{(c^0_{l3}+c^0_{e3}-1)}$.
 The lighter fermion masses are 
\bea
m_{f_2} &\sim&  \delta Y_Y\ m^0_{f1}\  \epsilon^{(\delta c_{L_2}+ \delta c_{R_2})},\\
m_{f1} &\sim& \delta Y_Y\ m^0_{f1}\  \epsilon^{(\delta c_{L_1}+ \delta c_{R_1})},\ \ \ \
\eea
where $m_{f2}\equiv m_c,m_s,m_\mu$, $m_{f1}\equiv m_u,m_d,m_e$ and
$m^0_{f1}= v\ \epsilon^{(c^0_{L1}+c^0_{R1}-1)}$. 
Note that in the flavor symmetric
limit, the electron and muon, the down and strange quarks, and the up and charm
quarks, are massless. The symmetry breaking
produces non-zero masses proportional to the generic size of $\delta Y$
among these fermions\footnote{The value of the light quark wavefunctions on the
TeV brane is slightly higher than in usual RS scenarios
to overcome the  $\delta Y$ suppression. This induces stronger couplings with KK gluons
and thus generically enhance dangerous FCNC processes.
These dangerous effects can be addressed by invoking additional flavor
constrains such as enforcing exactly $c_{d1}=c_{d2}$
\cite{Santiago:2008vq}, or going to more general warped backgrounds
where flavor bounds can be much milder \cite{Cabrer}.}, with
an added source of hierarchy due to the exponential
dependence on the $\delta c_i$. It is quite
simple in this scenario to obtain phenomenologically viable quark and lepton masses by
appropriately fixing the different $\delta c_i$ within the constraint
$|\delta c_i| \lsim 0.1$.
The hierarchies between fermion masses
occur naturally and are under control since they depend
exponentially on {\it small} numbers (they are hierarchical but not
too hierarchical).
Some masses and mixings still depend linearly on  $\delta Y$ 
so that the typical size of these terms cannot be too small since, for instance,
 $\delta Y \gsim m_t/m_c$ in the (extreme) limit
where the charm quark c-parameters are top-like.

The observed mixing angles in the CKM and PMNS matrices can also
be generated in this unified scenario. The CKM entries
become\footnote{$V_{cb}$ and $V_{ub}$ receive an extra
parametric suppression linear in the small Yukawa perturbations $\delta
  Y$, caused by the broken $S_3$ symmetry.}
 \bea
&& V_{us}\sim
\epsilon^{(\delta c_{q1}-
  \delta c_{q2})},  \\
&& V_{cb} \sim \delta Y\ \epsilon^{(\delta c_{q2}- \delta c_{q3})} \\
&&  V_{ub}\sim \delta Y\  \epsilon^{(\delta c_{q1}-
  \delta c_{q3})}\,.
\eea
 With respect to the 0-th order CKM matrix from
Eq.~(\ref{zeroorder}), $V_{us}$ receives a suppression
exponentially sensitive to the difference between two small terms with respect to the CKM,
which can easily reproduce the Cabibbo angle.
The angles $V_{cb}$ and $V_{ub}$, lifted from the initial zero value, 
acquire a double suppression, one exponential, and one proportional to
$\delta Y\sim 0.1$, with the ratio $V_{cb}/V_{us}$
of the correct order of magnitude for the 
typical size for $\delta Y$ and $\delta c$, assuming an ordering $\delta
c_{q1}>\delta c_{q2} >\delta c_{q3}$. 
The expected 
order of the ratio $V_{ub}/V_{cb}\sim V_{us}$ also remains realistic,
up to order one factors not taken into account in the estimates. 
This last feature is generic in usual RS scenarios. 

The parametric dependence of the PMNS entries
is different: 
\bea
&&V_{e2}\sim  \sin{\theta}^0_{\nu} \\
 && V_{e3}\sim \delta Y^\nu_{13}\ \sqrt{|\epsilon^{(2c_{l3}-1)}-1|} \label{ve3}\\
&&V_{\mu3}\sim \delta Y^\nu_{23}\ \sqrt{|\epsilon^{(2c_{l3}-1)}-1|}\, . \label{vmu3}\ \ \
\eea
Contrary to the quark sector, the value of $V_{e2}$ is not suppressed and remains
generically of ${\cal O}(1)$, fixed by the structure of the neutrino
Yukawa flavor violating matrix $\delta Y^\nu_{ij}$.
The entries $V_{e3}$ and $V_{\mu3}$ are lifted from zero, both
depend on $\delta Y$ and, not only are they not further suppressed by
exponential terms,  
but can actually be enhanced by exponential terms
(as long as the approximation remains valid). In particular if
$c_{l3}\lesssim 1/2$,  it is possible to lift the values of the 
mixing angles as shown in Eqs.~(\ref{ve3}) and (\ref{vmu3}). This 
feature is
specific to the case $a<c_{l_3}+c_{\nu_j}$ and $c_{l_3}<1/2$, and is
not generic in usual RS scenarios. 
More precise (and less compact) formulae will be presented in a companion
long paper.

The observed mixing angles in the neutrino sector are most sensitive to the flavor
structure of the neutrino Yukawa matrix $\delta Y^\nu$, but not  much
to the charged lepton Yukawa matrix $\delta Y^l$ or to the $\delta c_i$. 
The bulk mass parameter of the third
family lepton doublet should satisfy $c_{l_3}<1/2$  to easily obtain
larger mixing angles for small $\delta Y\sim 0.1$ (given that $V_{\mu3}^{\rm exp}\sim 0.65$ and
$V_{e3}^{\rm exp}\sim 0.15$).
This condition is very interesting as it is the same in the quark sector, where  
$c_{q_3} \lesssim 1/2$,  
to obtain a large top quark mass, which could be a hint of an additional family symmetry
among the $SU(2)$ doublets of the third family.
Comparing expressions for the $V_{PMNS}$ mixing angles and the
neutrino masses, the element $\delta Y_{23}$ must be
larger that the rest of $\delta Y$ so that $\delta Y \sim \delta Y_{13}
\sim \frac{\delta Y_{23}}{4}$. 

In this scenario it is easy to find a set of 0-th order bulk parameters that reproduce
 the SM and that show the features described
above.  For example,
a working point for which the SM is a small perturbation (of order $10\%$ around an $S_3$
symmetric set of parameters) away is shown in Table \ref{table1}.
\begin{table}[h]
\begin{center}
\begin{tabular}{c|c c c c c c} 
$f $& $q $&$ u$ & $d$ & $l $& $\nu$ & $e $\\
\hline\hline
$c^0_{f1}(\equiv c^0_{f2})$ & 0.55 & 0.60 & 0.60 & 0.55 & 5.00 & 0.60\\
$c^0_{f3}$ & 0.40 & 0.40 & 0.50 & 0.40 & 2.00 & 0.60
\end{tabular}
\end{center}
\vspace{-.4cm}
\caption{Zeroth order 5D fermion $c$-parameters. For simplicity, we also set all the 0-th order Yukawa coefficients to
be universal $y^0_u=y^0_d=y^0_\nu=y^0_e=4.4$ and the Higgs
localization parameter to $a=1.9$. \label{table1}}
\end{table}
Charged fermions results are not
too sensitive to small deviations in the Yukawa couplings ($\lesssim 0.1$), and once the $\delta c_i$'s  are fixed, the $\delta
Y$'s can even be taken randomly  as long as they remain at around
$10\%$. One then obtains generically charged fermion masses
and mixings consistent with the SM and any level of precision is possible
by tuning these values.

In conclusion, we have proposed a general framework in warped extra dimensions
where the SM flavor structure is unified in all fermion sectors. 
Small breaking terms are introduced for the 5D bulk mass and Yukawa
parameters.  The quark and charged lepton sectors are dominated by the small flavor 
breaking in the bulk $c$-parameters whereas the (Dirac) neutrino sector is
dominated by flavor symmetry breaking Yukawa couplings. The main difference between these stems
from allowing the Higgs field leak sufficiently out of the TeV
brane so that the neutrino sector looses sensitivity on the 5D bulk
mass parameters. 
A permutation  symmetry was studied to illustrate the idea, but 
other symmetries can be invoked and explored within this framework,
as will be further studied in a companion paper.

\acknowledgments
We thank NSERC for partial financial support  under grant number
SAP105354.


\end{document}